# Increasing the Number of Underrepresented Minorities in Astronomy at the Undergraduate, Graduate, and Postdoctoral Levels (Paper I)

*An Astro2010 State of the Profession Position Paper*
March 2009

**Authored by**: The AAS Committee on the Status of Minorities in Astronomy (CSMA),
**with endorsement from:**
National Society of Hispanics in Physics - David J. Ernst, Pres. (Vanderbilt University),
Marcel Agueros (Columbia University), Scott F. Anderson (University of Washington), Andrew Baker (Rutgers University), Adam Burgasser, (Massachusetts Institute of Technology), Kelle Cruz (Caltech), Eric Gawiser (Rutgers University), Anita Krishnamurthi (University of Maryland, College Park), Hyun-chul Lee (Washington State University), Kenneth Mighell (NOAO), Charles McGruder (Western Kentucky University), Dara Norman (NOAO), Philip J. Sakimoto (University of Notre Dame), Kartik Sheth (Spitzer Science Center), Dave Soderblom (STScI), Michael Strauss (Princeton University), Donald Walter (South Carolina State University), Andrew West (MIT)
UW Pre-Map staff - Eric Agol (Faculty Project Leader), Jeremiah Murphy, Sarah Garner, Jill Bellovary, Sarah Schmidt, Nick Cowan, Stephanie Gogarten, Adrienne Stilp, Charlotte Christensen, Eric Hilton, Daryl Haggard, Sarah Loebman Phil Rosenfield, Ferah Munshi (University of Washington)

Primary Contact
Dara Norman
NOAO
950 N. Cherry Ave
Tucson, AZ 85719
dnorman@noao.edu,
520-318-8361

*Increasing the Number of Underrepresented Minorities in Astronomy*
*at the Undergraduate, Graduate, and Postdoctoral Levels*

**ABSTRACT**

If the ethnic makeup of the astronomy profession is to achieve parity with the general population within one generation (~30 years), the number of underrepresented minorities earning graduate degrees in astronomy and astrophysics must increase in the coming decade by a factor of 5 to 10. To accomplish this, the profession must develop and invest in mechanisms to more effectively move individuals across critical educational junctures to the PhD and beyond. Early and continuous research engagement starting in the undergraduate years is critical to this vision, in which the federally funded research internship programs (e.g. NSF REU, NASA GSRP) and national centers/observatories play a vital role. Regionally based partnerships with minority-serving institutions (MSIs) are crucial for tapping extant pools of minority talent, as are post-baccalaurate and/or masters degree "bridging" programs that provide critical stepping stones to the PhD. Because of the strong undergraduate physics, engineering, and computer science backgrounds of many students from MSIs, we suggest that instrument development and large scale computing/data-mining are particularly promising avenues for engagement in the coming decade.

## 1. Statement of the Problem

The underrepresentation of minorities is one of the major challenges facing the US science and engineering workforce as a whole (see, e.g., the National Science Board's 2003 report, *The Science and Engineering Workforce: Realizing America's Potential*), and is more challenging still in the disciplines that sustain the astronomy and astrophysics enterprise[1]. Black, Hispanic, and Native Americans comprise 27% of the US population but are less than 4% of the astronomy workforce[2]. The nation's astronomy and astrophysics PhD-granting programs today collectively produce approximately 5±1 minority PhDs per year, an average per PhD-granting institution of 1 minority PhD every 10 years (Stassun 2005). Thus, at the same time that physical sciences PhD-granting programs in the US are increasingly turning to foreign students to make up the PhD ranks[3], these domestic pools remain largely underutilized. These same groups comprise 2.5% of physics and astronomy faculty at colleges and universities, and 0.5% of the physics and astronomy faculty at research institutions (statistics from AIP, Nelson & Lopez 2004).

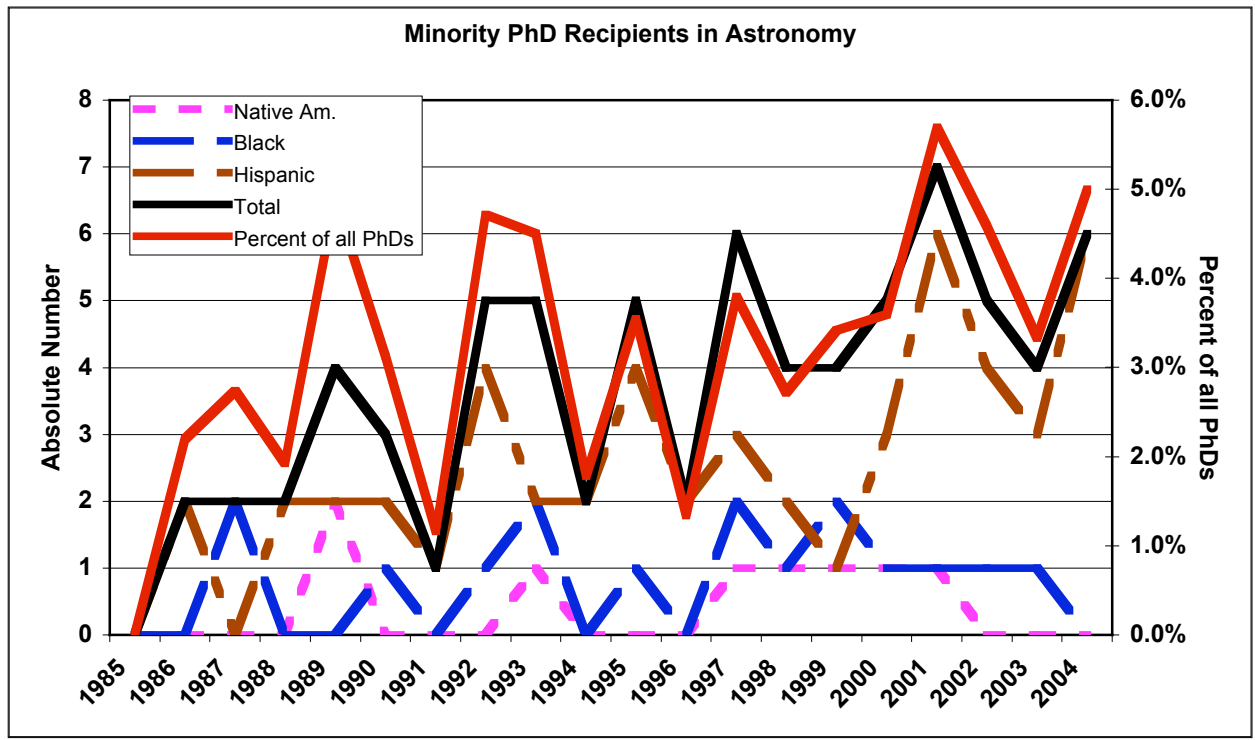

---

[1] By "astronomy enterprise" we mean Astronomy and related fields that include Physics, engineering and computer science. See the section on "Enhance recruitment through physics and engineering".

Unfortunately, the relative representation of these groups in astronomy and astrophysics has been steadily worsening. The fraction of minority PhDs has been roughly flat at 2-4% of the total (see figure above). However, during this same time period, the proportion of underrepresented minorities in the U.S. population grew by 33%, from 20.9% in 1988 to 27.0% in 2008 (data from US Census).

Estimates by the AAS Employment Committee find that on average 75±15 permanent astronomy jobs open up in the U.S. every year. Thus to achieve parity in the number of minorities entering the stream of permanent astronomy and astrophysics positions, the community must in the coming decade increase the number of minority PhDs from 5 per year to 20. And if the same attrition rates apply to these individuals as in the field overall, this number becomes 50. *Thus the absolute number of minority PhDs produced annually must increase by a factor of at least 5-10 in the coming decade.* At this aggressive pace, the field overall can achieve parity in 30-35 years.

It is moreover essential to realize that this scale of PhD production cannot be achieved by the current community of minority astronomers alone. A recent survey of all 50 astronomy PhD-granting programs counted a total of 17 individuals who identify as underrepresented minorities among the full-time faculty (Nelson & Lopez 2004). Increasing the number of minority PhDs to the levels required in the coming decade will have to be the purview of the entire astronomy and astrophysics community.

## 2. Solutions

Below we focus on practical solutions for increasing the participation of minorities in the Astronomy enterprise that can be achieved within the next decade. The solutions presented here can be implemented on short timescales because they tap into pools of students who are already at or near the required academic levels. These solutions can move us in the right direction quickly. At the same time, achieving fuller equity not only requires earnest and sustained commitments from academic/government institutions and funding agencies, but also a more active engagement in K-12 education from the astronomical community to see students through to these critical academic levels (see associated Position Paper on "Increasing the Number of Underrepresented Minorities in Astronomy through K-12 Education and Public Outreach (Paper II)".

**Develop meaningful partnerships with minority serving institutions (MSIs)**
MSIs (i.e. Historically Black Colleges and Universities, Hispanic Serving Institutions, Tribal Colleges) produce large pools of talent at the undergraduate level in the physical sciences and

---

[2] Throughout this report we adopt the definition of underrepresented minorities used by the federal funding agencies.
[3] In 2004, physical science PhD programs in the US awarded 9 times as many PhDs to international students than to domestic African American students (*Diversity and the PhD*, Woodrow Wilson National Fellowship Foundation, 2005).

| Universities that awarded the most physics bachelor's to African Americans. |
|---|
| Physics departments in these twenty universities awarded more than 55% of all physics bachelor's degrees earned by African Americans since 1998. |
| Alabama A&M University |
| Benedict College |
| Chicago State University |
| Delaware State University |
| Dillard University |
| Fisk University |
| Florida A&M University |
| Grambling University |
| Hampton University |
| Jackson State University |
| Lincoln University |
| Morehouse University |
| Morgan State University |
| Norfolk State University |
| North Carolina A&T State University |
| Southern University and A&M College |
| Spelman College |
| Tennessee State University |
| Tuskegee University |
| Xavier University |
| The physics departments on this list reported conferring 15 or more bachelor's degrees to African Americans between 1998 and 2007. |
| Source: AIP Statistical Research Center, Enrollment & Degrees Survey |

engineering. The top 15 producers of African American physics baccalaureates are all HBCUs. Moreover, these institutions are highly successful at placing students in PhD programs. For example, among the U.S. baccalaureate-origin institutions for Black science and engineering doctorate recipients, the top 8, and 20 of the top 50, were HBCUs (for 1997-2006; Burrelli & Rapoport, NSF, 2008). As we discuss below, partnerships with MSIs are a clear gateway to this talent pool, and to be effective such partnerships should be both "horizontal" and "vertical", and regionally based where possible.

*Horizontal partnerships*
Earnest partnerships should be sought between MSIs and major research institutions in which the MSIs are responsible for mission critical parts of a larger project and in which there are two-way exchanges of students and faculty. Such partnerships increase the capacity of the MSIs and—perhaps even more importantly—seed personal relationships that can lead to placement of MSI students in positions in the major institutions.

Examples of such partnerships supported through the (now defunct) NASA MUCERPI program are given in Sakimoto & Rosendhal (2005). Partnering of institutions should involve strong, authentic intellectual linkages that are mutually beneficial and that innovate around the synergies of the partnering institutions, building on complementary strengths (e.g. an astro-computational program pairing with an MSI with a strong computer science program).

*Vertical partnerships*
Recent research reveals that the master's degree is a critical transition point for many minorities at the graduate level in science and engineering. Master's education is a growing enterprise in U.S. colleges and universities. Much of that growth has been attributed to the entrance of groups underrepresented in graduate school enrollments – women and students of color. In the decade between 1990 and 2000, the total number of master's recipients increased by 42%. During this same time period, the number of women earning master's degrees increased by 56%, African Americans increased by 132%, Native American by 101%, and Hispanic by 146% (Syverson, 2003). Consequently, as shown in the figure below, underrepresented minorities in the physical sciences are ~50% more likely than their non-minority counterparts to earn a "terminal" master's degree (i.e. not a masters degree earned as part of a PhD program) before eventually transitioning to a PhD program (Lange et al. 2006). Thus too often, minority students must navigate multiple institutional transitions en route to the PhD, and the transition from undergraduate to master's to PhD is made with little or no structured mentorship (Stassun et al. 2008).

Explicit institutional/programmatic links and bridges should be built to attend to this and similar transitions in the career development pipeline, to further insure that potential PhDs are not lost. Promising example programs have recently emerged, including the Columbia post-baccalaureate "Bridge to PhD" program which links local college graduates to PhD programs at

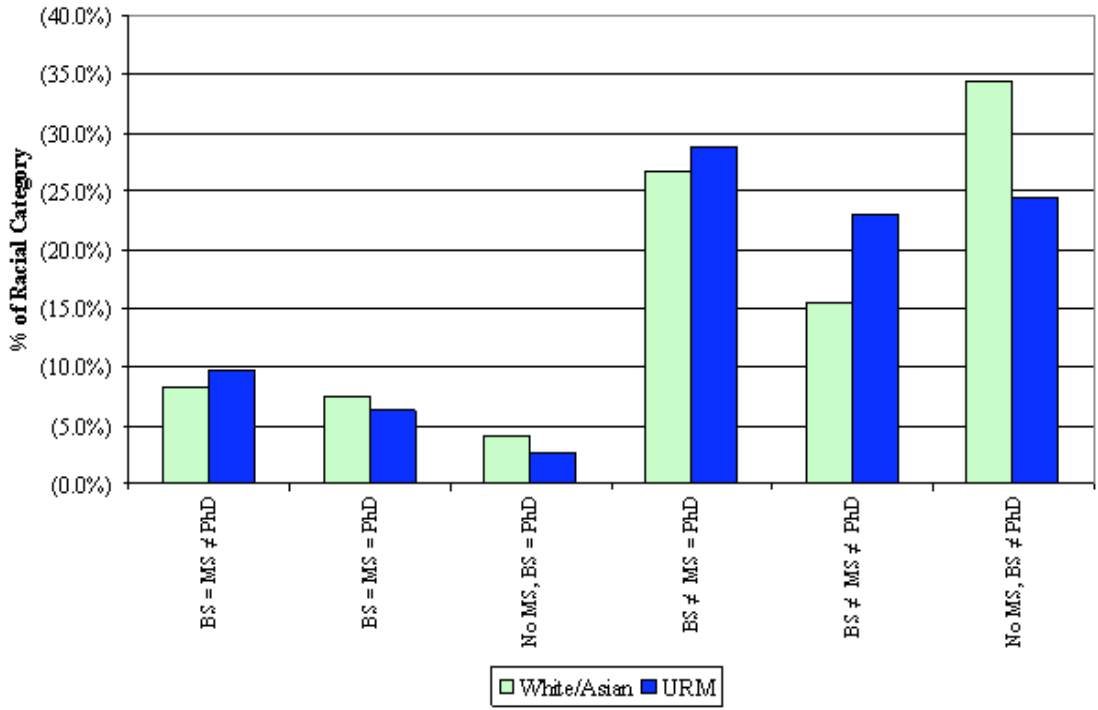

The graph shows comparisons between underrepresented minorities (URMs) and White/Asian students, based on different permutations of the educational pathway to the PhD. An equal sign indicates degrees earned from the same institution. The fourth and sixth comparisons from the left show the "traditional" paths to the PhD, in which the student earns the bachelors degree from institution A, and either receives both the masters degree and the PhD from institution B or else forgoes the masters degree entirely. The fifth comparison from the left is shown the case for earning the bachelors degree at institution A, a "terminal" masters degree at institution B, and PhD from institution C. Minorities are much more likely to take this latter path than non-minorities. Adapted from Lange (2006) based on analysis of 80,739 PhDs earned in science and engineering fields, 1998 to 2002.

Columbia or elsewhere through a two-year program of research internships and mentoring, and the Fisk-Vanderbilt Masters-to-PhD Bridge program (with funding from NASA MUCERPI and NSF PAARE) through which students use the MA degree at Fisk as a stepping stone to the PhD at Vanderbilt through collaborative research projects at the two institutions.
(Program details can be found at www.columbia.edu/cu/vpdi/bridge_students.htm and www.vanderbilt.edu/gradschool/bridge and in the Appendix[4].)

    As promising as these programs are, long-term sustainability remains a concern. Not all programs are institutionalized, and even those that do receive significant institutional support must still rely upon substantial federal funding (see, e.g., the Fisk-Vanderbilt Bridge or the Pre-MAP program financial support information, both in the Appendix). Funding for these programs from the federal agencies has in almost all cases been explicitly limited in duration (i.e. no competitive renewals). The requirement of "exit strategies" for successful and innovative programs is counter to the clear needs of the profession for the human capital that these programs produce.

---

[4] Appendices for this paper can be found on the AAS CSMA website at http://csma.aas.org/events.html

The funding agencies should take a more balanced approach toward seeding untested initiatives while also providing the opportunity for stable support of the most productive and innovative programs.

*Geographic partnerships*
Potential partners should not only be aware of, but also take advantage of, the strong geographic concentration of MSIs. HBCUs are predominantly clustered in the Southeast, HSIs in the Southwest, and TCs in the Northwest/Midwest (see maps and lists of MSIs in the Appendix). To the extent that minority students may in some cases be "location bound" for familial, cultural, or other reasons, regionally based partnerships may be particularly effective at successfully recruiting, retaining, and transitioning these students (e.g. Stassun 2003). Indeed, most of the partnerships discussed by Sakimoto & Rosendhal (2005), including those specifically mentioned in this report, have a regional basis. However, not all successful partnership programs are regional. For example, the NSF Center for Integrated Space Modeling links Alabama A&M and Boston University.

Finally, we note that partnership programs of the type discussed above need also to be developed to address the transition from the PhD to the workforce. Some successful programs in academia exist (e.g. the University of California Presidential Postdoctoral Fellows program). The National Centers and Observatories are in a position to play an important role here as well. For example, the NASA Co-op program identifies individuals early in their graduate careers and tracks them into permanent civil servant positions. Similar approaches could be explored by the National Observatories and other centers that support tenured positions.

**Enhance recruitment through physics and engineering as on-ramps to astronomy**
Of the ~100 HBCUs in the United States, only three currently offer a formal astronomy curriculum at the undergraduate level—an astronomy minor in all cases[5]. No HBCU offers an astronomy major. However nearly all of them offer the physics major, and many offer engineering majors including computer science. While a degree in physics has traditionally presented a clear path to an astronomy career, engineering skills are clearly needed to meet the future science goals of US Astronomy.

The development of next generation instrumentation for large telescopes and space-based missions will require technical expertise which includes engineering and design, systems management and proper administration of projects (Astronomy and Astrophysics in the New Millennium, NRC 2000). It will be advantageous to the astronomical enterprise to both engage traditionally trained engineers and to also train engineers to have expertise in astronomy. For example, the LSST FaST program (funded by NSF/DOE) involves faculty and student teams from MSIs in the design and development of LSST software and hardware.

The increasing importance of theoretical large scale computation for simulation, data-intensive surveys (such as LSST), and data-mining infrastructures (such as the Virtual Observatory) require computer savvy astronomers who might be tapped from the ranks of computer science graduates. Thus these more non-traditional majors represent promising avenues for engagement of students with these disciplinary backgrounds into the astronomical enterprise in relatively short order.

---

[5] Fisk University (Nashville, TN), Southern University (Baton Rouge, LA), South Carolina State University (Orangeburg, SC).

In parallel with direct outreach to faculty and students in these "on-ramp" disciplines at MSIs, the astronomical community should also support the hiring of faculty with astronomy-related interests in these same institutions' on-ramp departments. Whether a condensed matter experimentalist with an interest in detector development or a computational theorist specializing in compact objects, such individuals help expose students to astronomy opportunities (such as REU and graduate programs), help bring those students to the attention of the larger astronomical community, and serve as obvious points of contact for establishing horizontal and vertical partnerships as described above.

**Provide early and continuous engagement in research**

Studies have shown that the exposure to current research early in the careers of all students encourages them to continue on in the pursuit of a doctoral degree (Nadga 1998). As we seek to increase the numbers of minority students in astronomy, we must do a better job of including these students in current and ongoing research experiences. Although there are formal programs at some colleges and universities to involve undergraduates in current research with local faculty, institutions generally need to bring that knowledge to minority students. Astronomy departments must more actively recruit underrepresented students to participate in local research. This recruitment process is important whether done directly by departments or by institutional offices (like Offices of Minority Education) because students may not be aware that such research opportunities exist. An example of such a program is the University of Washington's Pre-Major in Astronomy Program (Pre-MAP), which is a targeted effort to introduce students from traditionally underrepresented groups to astronomy research as early as the first quarter of the freshman year. (See www.astro.washington.edu/users/premap/ and the Appendix for details.)

The NSF's Research Experiences for Undergraduates (REU) program, and other undergraduate internship programs like it, have been a successful way to engage undergraduate students in Astronomy research outside of their home institution. The program has helped to increase the number of women in astronomy (e.g. Russell 2005). The NSF must encourage their REU sites to recruit students from traditionally underrepresented groups and regional areas. In addition, programs like this should be expanded to institutions with large minority serving communities so that students from all backgrounds flow through these institutions and engage with professors and students in these institutions. At present, 2 of the 20 NSF REU sites in Astronomy are hosted at minority-serving institutions.

More generally, strong funding support for REU and similar research immersion programs is essential to maintaining the vitality of the higher education pipeline in astronomy. In addition, in order to help attend to the critical undergraduate-to-graduate transition point, the agencies should relax current restrictions against support for students who have completed the senior year of college but have not yet enrolled in a graduate program.

Research centers, national labs, and observatories play a critical role in this arena. The scientific staff at these facilities represent an extended pool of mentoring opportunities for students, especially those from smaller institutions where research-active astronomy mentors and access to research facilities may be in short supply. In addition, many of the national centers host groups and facilities that are active in the development of instrumentation and/or in data-intensive computing for the next generation of large surveys, areas which students from schools with strong physics/computer science/engineering programs may find particularly attractive (see above). For example, the national centers and observatories could initiate instrumentation internship programs, and large survey projects in which these centers are involved could provide internship opportunities in data-intensive astrophysics.

*Increasing the Number of Underrepresented Minorities in Astronomy*
*at the Undergraduate, Graduate, and Postdoctoral Levels*### 3. Specific recommendations

Recommendations for PhD granting institutions:
1. Form "horizontal" partnerships with MSIs in which MSI students and faculty are equal stake holders in mission critical research, funding, and development opportunities. Institutional match-ups should reflect authentic synergies of intellectual contribution. In addition, the strong geographic concentration of MSIs may make regionally based partnerships both feasible and more attractive to "location bound" students.
2. Develop "vertical" partnership programs that ease the transition of minorities across critical junctures in the pipeline (college freshman transition programs, post-baccalaureate programs, Masters-to-PhD bridging programs). The Masters degree is emerging as a critically important, and previously poorly understood, transition point.

Recommendations for funding agencies:
1. Substantially expand funding for undergraduate research internships. Extend REU support to students in the summer between the senior year of college and the first year of graduate school. Alternatively, develop opportunities for funding of "post-baccalaureate" students in research training programs.
2. Establish and maintain a suitable number of REU site programs at MSIs. These programs can bring in students from both minority and majority institutions to advance development of horizontal partnerships.
3. Provide new, focused support for programs and research portfolios that serve as on-ramps to astronomy from physics, engineering, and computer science. For example, the development of next generation instrumentation for large telescopes and space-based missions, and the increasing importance of large scale computation for simulation and data-mining, both represent promising avenues for engagement.
4. Substantially expand funding for programs that specifically forge linkages between minority-serving institutions and research universities (e.g. NSF PAARE, NSF PREM, NASA MUCERPI).
5. Provide funding incentives for broadening participation of underrepresented minorities in federally funded programs by including this in funding criteria (e.g. NSF's "broader impacts" criterion).
6. Provide opportunities for continuity of funding: Federal funding for minority engagement can achieve continuity of the most successful programs by establishing funding cycles in the same way that research is funded. The current practice of awarding seed funding with restrictions against innovative renewals is deleterious to the success and long-term viability of these programs.
7. Collect and maintain reliable statistics on minority representation and participation. There have been recent calls to limit collection of certain statistics because of the small numbers involved. However, these small numbers appear in demographics that are in areas crucial for tracking progress.

Recommendations for National Centers, Labs, Observatories:
1. Develop internship programs that connect minority students to mentored research engagement with scientific and/or engineering staff.
2. Develop pipeline programs in partnership with universities. Especially needed are pipelines to post-doctoral opportunities.

Recommendations for professional societies:
1. Create a professional network to better link potential candidates (at all levels) with potential employers/programs. The American Mathematical Society has recently created such a network that could be used as a model.
2. Enhance connectivity and networking between the traditional professional societies (e.g. AAS) and those that principally serve and represent minority students and professionals (e.g. National Society of Black Physicists, National Society of Hispanic Physicists, Society for the Advancement of Chicanos and Native Americans in Science, Society of Hispanic Professional Engineers).

*Increasing the Number of Underrepresented Minorities in Astronomy
at the Undergraduate, Graduate, and Postdoctoral Levels*

# APPENDICES

## Geographical Distribution of Minority Serving Institutions

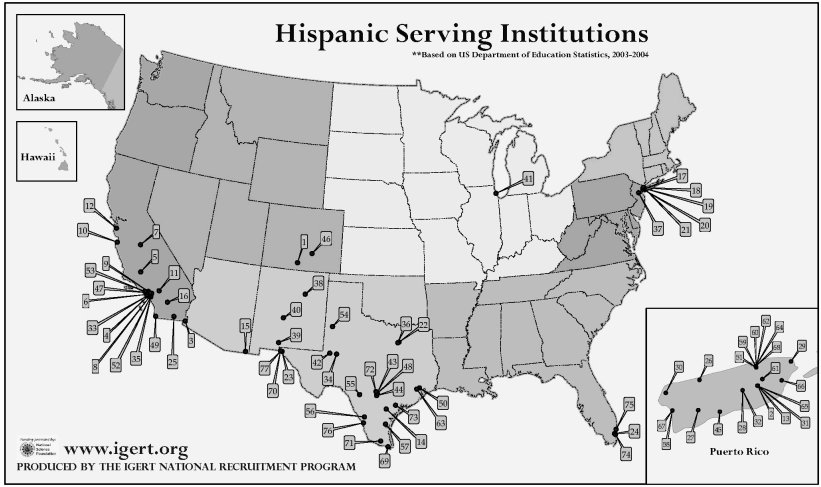

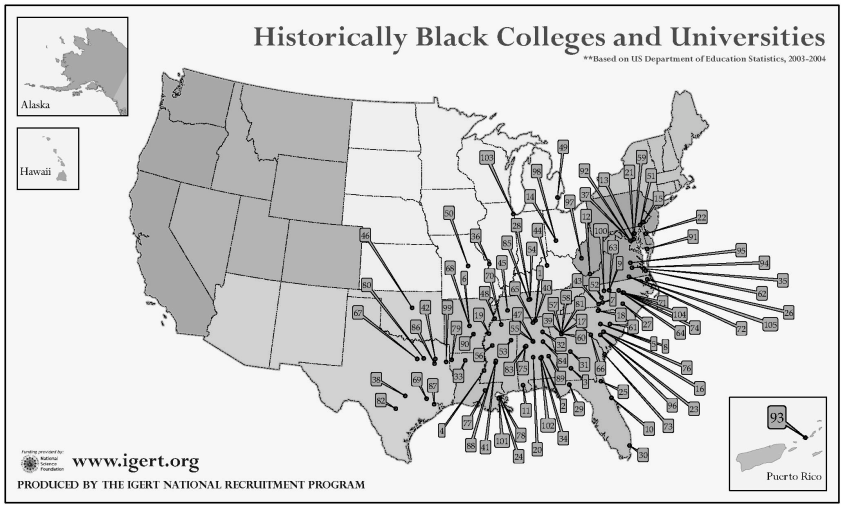

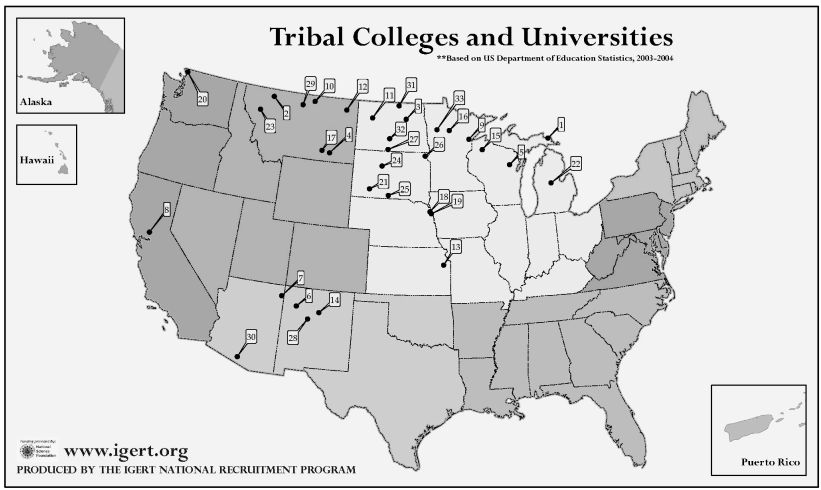

See also the Jan 2006 issue of the Spectrum newsletter for comprehensive lists of these MSIs.

## Colleges/Universities with Physics Departments/Bachelors in Physics – Southeast Region
**(Those with Masters and PhDs as indicated)     (Chairs)**

## Last updated: 05/24/05

### HBCU's

AL
Ala A&M plus Masters (M) & PhD    Aggarwal, M.D.      mohan.aggarwal@email.aamu.edu
Tuskegee                                              Sharma, Prakash     pcsharma@tuskegee.edu

AR: Univ of Ark at Pine Bluff                  Iyere, Peter         iyere@uapb.edu

DE: Delaware State plus M                      Helmy, Ehsan         ehelmy@desu.edu

DC: Howard plus M and PhD                      Venable, Demertius   dvenable@howard.edu

FL
Fla A&M plus M and PhD                         Mochena, Mogus       mogus.mochena@famu.edu

GA
Clark Atlanta plus M                           Williams, Michael    mdwms@cau.edu
Morehouse                                      Dixon, Robert        rdixon@morehouse.edu
Spelman                                        Hylton, Derrick      dhylton@spelman.edu

KY: none

LA
Grambling                                      Ware, Michael F.     waremf@gram.edu
Southern at Baton Rouge plus M                 McGuire, Stephen     mcguire@phys.subr.edu
Xavier of LA                                   McCloud, Kathleen    kmccloud@xula.edu

MD: Morgan State                               Oliver, Frederick    fwoliver@morgan.edu

MS
Jackson State                                  Williams, Quinton L. quinton.l.williams@jsums.edu
Tugaloo College                                Banerjee, Santanu    sbanerjee@tougaloo.edu

NC
NC A&T plus M                                  Bililign, Solomon
NC Central                                     Vlahovic, Branislav  vlahovic@nccu.edu

OK: Langston                                   Snow, Joel           snow@physics.lunet.edu

SC
Benedict                                       Arammash, Fouzi      farramash@benedict.edu
SC State                                       Salley-Guydon, Judicth djdsalley@scsu.edu

TN
Fisk plus M                          Morgan, Steven         smorgan@fisk.edu
Tenn State                           Scheick, Sandra        sscheick@tnstate.edu
TX: Prairie View                     Kumar, A. Anil         anil_kumar@pvamu.edu

VA
Hampton plus M and PhD               Temple, Doyle          doyle.temple@hamptonu.edu
Norfolk State plus M (materials sci) Ferguson, Milton       mvfergusin@nsu.edu
Va State plus M                      Gatrone, Ralph         rgatrone@vsu.edu

WVA: West Va State College

# HSI's

FL
Fla International plus M and PhD     VanHamme, Walter    walter.vanhamme@fiu.edu
U of Miami plus M and PhD            Alexandrakis, G    alexandrakis@physics.miami.edu

NM: New Mexico State-Las Cruces      Kyle, Gary             gkyle@nmsu.edu

TX
St. Mary's Univ                      Cardenas. Richard      rcardenas@stmarytx.edu
TX A&M Kingsville                    Suson, Daniel          daniel.suson@tamuk.edu
TX State-San Marcos plus M&PhD Gutierrez, Carlos-contact  cgutierrez@txstate.edu
U of Houston plus M and PhD          Pinsky. Lawrence       pinsky@uh.edu
UT Brownsville plus M                Guevara, Natalie       nguevara@utb.edu
UT El Paso plus M                    Lopez, George          jorgelopez@utep.edu
UT San Antionio plus M and PhD       Nash, Patrick          patrick.nash@utsa.edu
UT Permian Basin                     Allen, Donald          allen_d@utpb.edu
UT Pan Am                            Mazariegos, Ruben      rubenm@panam.edu

Puerto Rico
Pontifical Catholic U-Ponce          Perez, Lucinda         lperez@email.pucpr.edu
U of PR Rio Piedras plus M and
 PhD (chemical physics)              Fonseca, Luis          chairman@physd.cnnet.clu.edu
U of PR Mayaguez plus M
U of PR Humacao                      Cersosimo, Juan Carlos j_cersosimo@webmail.uprh.edu
U Metropolitana-San Juan
  Model Institution for Excellence   Arratia, Juan          um_jarratia@suagm.edu

Pre-MAP Press Kit

premap@astro.washington.edu
http://www.astro.washington.edu/premap/

## Program Overview

The Pre-Major in Astronomy Program (Pre-MAP) is a research and mentoring program for first year students offered by the University of Washington Astronomy Department. The primary goal of Pre-MAP is to increase retention of students traditionally underrepresented in science, math and technical majors. Eligible students enroll in the Pre-MAP seminar to learn astronomical research techniques that they apply to research projects conducted in small groups. Students also receive one-on-one mentoring and peer support for the duration of the academic year and beyond. Additionally, they are invited to attend Astronomy Department events and Pre-MAP field trips. Successful Pre-MAP students have declared Astronomy and Physics majors, expanded their research projects beyond the Autumn quarter, presented posters at the UW Undergraduate Research Symposium (UGRS), and received both research fellowships and summer internships.

### Motivation
Currently the percentage of science Ph.D.s awarded to women, African American, Latino, and other minority students is far smaller than the percentage these groups constitute of the general population. The greatest obstacles for persistence in studying science reported by students are loss of interest, intimidation, poor advising, and lack of acceptance in their department. To address these concerns, Pre-MAP was developed to build a sense of community; to provide advising, mentoring, and academic support; and to engage students in discipline-specific research.

### Strategy
Pre-MAP develops a learning community of students interested in scientific research by offering a seminar for up to 15 participants, to be taken in conjunction with Astronomy 102, an introductory course offered Autumn quarter. The three-credit Pre-MAP seminar meets three hours a week and offers instruction in basic computing skills, data manipulation, science writing, and statistical analysis. Students choose from research projects presented by Astronomy faculty, post-doctoral fellows, and graduate students and are encouraged to continue their research during Winter and Spring quarters for academic credit. The seminar leader meets with students individually and mentors them formally during Autumn quarter. Mentoring and cohort building activities such as field trips, study sessions and outreach events continue throughout the remainder of the students' UW careers. Recruitment for the program begins before Autumn quarter through referrals, outreach events and self-enrollment.

### Outcomes
In October 2005, the first cohort of seven Pre-MAP students entered UW. In the following years the program has steadily grown, with fifteen students currently participating in Pre-MAP's fourth cohort. All groups of participating freshmen reported an increased interest in astronomy and science careers at the end of the Autumn quarter. Students participated in research ranging from searching for supernovae to using low mass stars to probe the Milky Way's structure. Students from cohorts one and two submitted almost half a million pairs of observations of asteroids from the SDSS-II supernova survey to the Minor Planet Center. The Center is currently reviewing this list, with the preliminary results that Pre-MAP students have added observations to the orbits of 13,487 previously known asteroids and have also discovered 850 new asteroids. As one mentor noted, "There are about 335,000 total known objects in the solar system. So [Pre-MAP students] actually affected the orbits of 4% of the known solar system objects!"

The majority of students have continued research throughout Winter and Spring quarters. Six students from the first cohort presented their findings at the university-wide Undergraduate Research Symposium (UGRS) in 2006; ten students followed suit and presented their research at the 2007

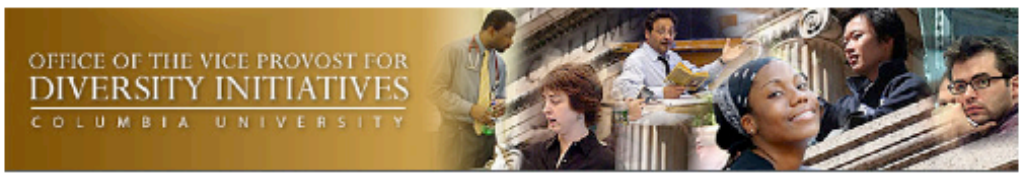

## The Columbia University Bridge to Ph.D. Program in the Natural Sciences

The Bridge to the Ph.D. Program aims to provide an intensive research, coursework, and mentoring experience to talented post-baccalaureate students from ethnic groups that have been historically underrepresented in the natural sciences. As a pipeline-building initiative, the Bridge Program seeks to enhance students' candidacy for acceptance to Ph.D. programs.

Bridge students are provided with full-time, salaried positions as Columbia University Research Assistants for up to two years. This opportunity provides exposure to laboratory-based science under the supervision and mentorship of faculty members, post-doctoral researchers, and graduate students. The current salary for Research Assistant (RA) positions is $32,300 per annum. Program participants are also provided with discretionary funding of $1500 per year to support professional and educational expenses (examples include travel to professional conferences or graduate school interviews, application or course registration fees, or the purchasing of books or journal subscriptions). As full-time employees of Columbia University, RAs are eligible for University benefits including medical insurance, transit reimbursement, and retirement benefits, among others.

Additionally, Bridge students enroll in one course per semester at Columbia that is related to their future field of study and attend monthly one-on-one progress meetings with a program advisor. Students do not pay tuition for these courses, but are responsible for registration and course fees. Through Columbia's School of Continuing Education, program participants also have access to GRE test preparation, writing workshops, time management courses, and other services designed to facilitate the process of applying to graduate programs.

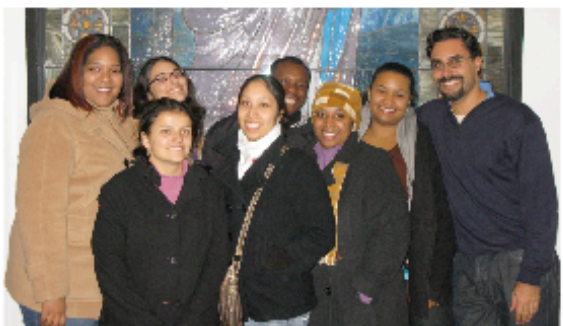

*The current cohort (missing: Vanessa Mondol).*

**Current Cohort**

| Student | Undergraduate Institution | Discipline |
|---|---|---|
| Ximena Fernandez | Vassar College | Astronomy |
| Tashina Graves | Barnard College | Psychology |
| Charlotte Logan | University of Oklahoma | Biology |
| Vanessa Mondol | SUNY Stony Brook | Chemistry |
| Chukwudi Onyemekwu | Vanderbilt University | Psychology |
| Elizabeth Rodriguez | Hunter College | Psychology |
| Nitza Santiago-Figueroa | University of Puerto Rico, Humacao | Astronomy |
| Marlena Watson | Temple University | Environmental Science |

# The Fisk-Vanderbilt Masters-to-PhD Bridge Program
www.vanderbilt.edu/gradschool/bridge or www.fisk.edu/bridge

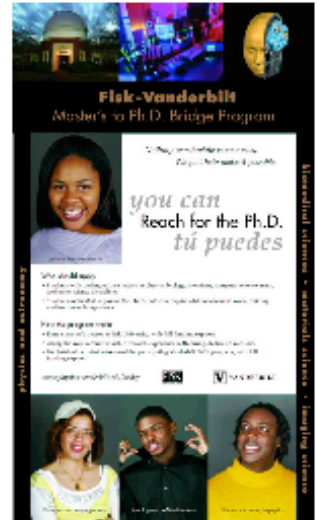

### Bridge Program Overview

The aim of the *Fisk-Vanderbilt Masters-to-PhD Bridge program* is to give students with undergraduate science or engineering degrees the preparation needed to earn a PhD in astronomy, physics, materials science, or biology. By completing a MS degree at Fisk under the guidance of caring faculty mentors, students develop the strong academic foundation, research skills, and one-on-one mentoring relationships that will foster a successful transition to the PhD at Vanderbilt. The program is flexible and individualized to the goals and needs of each student. Courses are selected to address gaps in undergraduate preparation, and research experiences are provided that allow students to develop—and to demonstrate—their full scientific talent and potential. Since 2004, the program has attracted 32 students, 29 of them underrepresented minorities[1] (59% female), with a retention rate of 94% (see Table 1). When the current *Bridge* students begin completing their PhD degrees in 2010-11, Vanderbilt will achieve the distinction of becoming the top research university to award PhDs to minorities in physics, astronomy, and materials science. Already, as of 2006 no U.S. institution awards more MS degrees in physics to Black U.S. citizens than Fisk.

### Bridge Program Facts & Figures

- In 2006, U.S. institutions awarded to Black U.S. citizens just 12 PhDs in physics (out of 637 U.S. citizen PhDs; 1.9%) [data from NSF WebCASPAR]. The average per PhD-granting institution in the U.S. is 1 minority[1] PhD in biology, physics, materials science, and astronomy every 2, 5, 9, and 13 years, respectively.
- The *Fisk-Vanderbilt Bridge program* is on track to award *10 times* the U.S. institutional average number of minority PhD recipients in astronomy, *9 times* the average in materials science, *5 times* the average in physics, and 2 times the average in biology (the biology track was newly added in 2007). Our most recent incoming cohort alone includes more minority students in astronomy than the current annual production of minority PhD astronomers for the entire U.S.
- Our *Bridge* students have been awarded the nation's top graduate fellowships from NSF (GRF and IGERT) and NASA (see Table 1 below).
- Extramural grants received to support the *Bridge* program—support for graduate students, faculty, and related undergraduate research—now exceed $10.8M (see Table 2). NSF and NASA research grants to support Bridge students and faculty totaling ~$14M are currently pending.
- Vanderbilt and Fisk now provide significant institutional support in the form of tuition waivers, RA stipends, and administrative support[2] (see Table 2).

### Table 1: Fisk-Vanderbilt Masters-to-PhD Bridge Program Students to Date

| Student | Ethnicity/Gender* | Admit Year | Undergraduate Institution | Discipline | Current Institution / Status |
|---|---|---|---|---|---|
| T. LeBlanc | H/M | 2004 | UMET, Puerto Rico | Astronomy | Vanderbilt (NASA Fellow) |
| J. Harrison | A/M | 2004 | Chicago State Univ. | Materials | Case Western (IGERT fellow) |
| H. Jackson | A/F | 2004 | Fisk University | Physics | Air Force Research Lab |
| J. Rigueur | A/M | 2004 | Fisk University | Physics | Vanderbilt (IGERT fellow) |
| V. Alexander | A/M | 2005 | Florida A&M Univ. | Physics | Dropped out, status unknown |
| J. Bodnarik | W/F | 2005 | USAF Academy | Astronomy | Vanderbilt (NASA Co-op) |
| M. Harrison | A/F | 2005 | Xavier University | Materials | Vanderbilt (IGERT fellow) |
| J. Isler | A/F | 2005 | Norfolk State Univ. | Astronomy | Yale (NSF graduate fellow) |

---

[1] Black-, Hispanic-, and Native-Americans comprise roughly 30% of the U.S. population.
[2] A full-time program coordinator is provided at Vanderbilt. Similar administrative support at Fisk remains a need.